\documentclass[prd,onecolum,preprintnumbers,showpacs,amsmath,amssymb,eqsecnum,nofootinbib]{revtex4}
\usepackage{mathpazo}
\usepackage{graphicx}
\usepackage[dvipdfm]{hyperref}
\usepackage{dcolumn}
\usepackage{bm}
\preprint{DPNU-06-09}
\begin{document}
\title{Breakdown of self-similar evolution in homogeneous perfect fluid collapse}

\author{Eiji Mitsuda}
\email{emitsuda@gravity.phys.nagoya-u.ac.jp}
\author{Akira Tomimatsu}
\email{atomi@gravity.phys.nagoya-u.ac.jp}
\affiliation{Department of Physics, Graduate School of Science, Nagoya
University, Chikusa, Nagoya 464-8602, Japan}
\begin{abstract}

The stability analysis of self-similar solutions is
 an important approach to confirm whether they act
 as an attractor in general non-self-similar gravitational
 collapse. Assuming that the collapsing matter is a perfect fluid with
 the equation of state $P=\alpha\rho$, we
 study spherically symmetric non-self-similar perturbations in
 homogeneous self-similar collapse described by the flat Friedmann solution.
In the low pressure approximation $\alpha \ll 1$, we analytically derive 
an infinite set of the normal modes and their growth (or decay) rate. 
The existence of one unstable normal mode is found to conclude that the
 self-similar behavior in homogeneous collapse of a
 sufficiently low pressure perfect fluid must terminate and a certain
 inhomogeneous density profile can develop with the lapse of time.

\end{abstract}
\maketitle

\section{Introduction\label{sec:intro}}

Spherically symmetric self-similar gravitational collapse of a perfect fluid with
pressure $P$ given by the equation of state $P=\alpha\rho$ is one of the
most extensively-studied phenomena in general relativity.
Many efforts have been made to solve the Einstein's
equations governing its dynamics, which are reduced to a
set of ordinary differential equations with respect to the single
variable $z\equiv r/t$. 
The flat Friedmann solution is well known as the unique analytically-found
exact solution regular at the center and has played
an important role in finding a family of solutions regular
at the center \cite{BJCarr:PRD62:2000, CarrBJ:CQG16:1999}. 
The homogeneous collapse described by this solution has been
considered as the most basic process to spacelike singularity formation in the self-similar dynamics, while 
essential features of inhomogeneous collapse 
have been understood mainly through the detailed analysis
of the general relativistic Larson-Penston solution \cite{OriA:PRD42:1990}.
In addition, it is also noteworthy that the perfect fluid critical collapse
corresponding to the threshold of black hole formation has been
confirmed to be described by one of the self-similar solutions \cite{EvansCR:PRL72:1994, CarrBJ:PRD61:2000, NeilsenDW:CQG17:2000}.

Several works have been devoted to the stability analysis of such
self-similar solutions for spherically symmetric non-self-similar
perturbations and have given important implications to more general
non-self-similar gravitational collapse. In particular, it is remarkable
that 
the flat Friedmann solution and the general relativistic
Larson-Penston solution were numerically confirmed to be able to act as an attractor
in general spherically symmetric gravitational collapse for $\alpha$
lying in the range $0<\alpha\lesssim 0.036$ \cite{HaradaT:PRD63:2001}.
In addition, the critical phenomena are 
illustrated in terms of the time evolution of the single unstable normal
mode which was found by the numerical analysis of the perturbations in the critical collapse
\cite{KoikeT:PRL74:1995, KoikeT:PRD59:1999}. (See
\cite{CarrBJ:GRG37:2005} for a recent review on the role of self-similar solutions as an
attractor and the critical phenomena.) Although it may be interesting
to study the stability problem more extensively, 
these numerical results should be confirmed 
through an analytical treatment of the perturbations.

Recently, we have developed an analytical scheme to treat the stability
problem by constructing the single wave equation governing
non-self-similar spherically symmetric perturbations \cite{MitsudaE:PRD73:2006}, 
which is reduced to the ordinary differential equation if we assume the
perturbations to have the time dependence given by
$\exp{(i\omega\log|t|)}$.
In this paper, using this analytical scheme, 
we study the stability problem for the flat Friedmann
solution in the low pressure limit, i.e., $0<\alpha\ll 1$. 
Fortunately, in the expansion with respect to the small parameter $\alpha$, we can explicitly solve the master ordinary differential equation for
the normal modes and consequently find the single unstable normal mode, which was not
found in the numerical analysis \cite{HaradaT:PRD63:2001}. 

We begin with a brief description of the perturbation theory 
for the flat Friedmann solution in Sec.~\ref{sec:theory}.
In Sec.~\ref{sec:normal}, a discrete set of the normal modes
and their growth (or decay) rate are derived in the low pressure limit, and the
self-similar behavior turns out to be unstable in homogeneous collapse of 
a sufficiently low pressure perfect fluid.
In Sec.~\ref{sec:interpretation},  
we see density inhomogeneities generated by the normal modes and
explain the affection of the background transonic flow
upon the growth (or decay) of the normal modes. 
In the final section, 
we summarize this paper and give a suggestion for the result of the
numerical study \cite{HaradaT:PRD63:2001}.
In addition, we discuss an implication of the breakdown of the
self-similar evolution in relation to critical phenomena.

\section{Perturbation theory for self-similar homogeneous perfect fluid
 collapse\label{sec:theory}}

In this section, 
we briefly illustrate our analytical scheme to treat spherically
symmetric non-self-similar perturbations in homogeneous self-similar
perfect fluid collapse described by the flat Friedmann solution (see
\cite{MitsudaE:PRD73:2006} for the details).
The line element considered throughout this paper is given by
\begin{equation}
 ds^{2}=
  -e^{2\nu(t,r)}dt^{2}+e^{2\lambda(t,r)}dr^{2}+R^{2}(t,r)\left(d\theta^{2}+\sin^{2}\theta d\varphi^{2}\right)\label{eq:LE}
\end{equation}
with the comoving coordinates $t$ and $r$.
In addition, the energy momentum tensor of a perfect fluid is
expressed as 
\begin{equation}
 T^{ab}=(\rho+P)u^{a}u^{b}+Pg^{ab}~,\label{eq:EM}
\end{equation}
where $\rho$, $P$ and the vector $u^{a}$ are energy
density, pressure and fluid four velocity, respectively.
As was mentioned in Sec.~\ref{sec:intro}, 
its equation of state is assumed to be 
\begin{equation}
 P = \alpha \rho \label{eq:EOS}
\end{equation}
with a constant $\alpha$ lying in the range $0<\alpha\leq 1$. 
To discuss the self-similar behavior later, we use a new variable $z$
defined by $z\equiv r/t$, instead of $r$.
In addition, instead of the four unknown functions $\nu$, $\lambda$, $R$
and $\rho$,  we introduce 
the following dimensionless functions:
\begin{equation}
 S(t,r) \equiv \frac{R}{r}~, \qquad \eta(t,r)\equiv 8\pi r^{2}\rho~,
  \qquad M (t,r) \equiv \frac{2m}{r}~, \qquad V(t,r) \equiv
  ze^{\lambda-\nu}~,
\end{equation}
where the function $m(t,r)$ is the Misner-Sharp mass.
The function $V$ is interpreted as the velocity of a $z=\text{const}$ surface
relative to the fluid element.

From the Einstein's field equations for
the system (\ref{eq:LE}), (\ref{eq:EM}) and (\ref{eq:EOS}), 
we can obtain the four partial differential
equations governing the functions $\nu$, $\lambda$, $S$ and $\eta$.
By the two equations, the metrics $\nu$ and $\lambda$ are explicitly
given by the functions $S$ and $\eta$.
The relations between the functions $S$ and $\eta$ given by the
remaining two equations become the simpler first order partial
differential equations if we use the function $M$, which is explicitly
given by the functions $S$ and $\eta$. Therefore we hereafter focus our concern on the 
functions $S$, $\eta$ and $M$.

Now we consider spherically symmetric non-self-similar perturbations in 
the flat Friedmann background by expressing
 the solutions for the Einstein's equations as 
\begin{eqnarray}
 &&S(t,z)=S_{B}(z)\left\{1+\epsilon S_{1}(t,z)+O(\epsilon^{2})\right\}~,
  \qquad
  \eta(t,z)=\eta_{B}(z)\left\{1+\epsilon\eta_{1}(t,z)+O(\epsilon^{2})\right\}~,\nonumber\\ 
 &&M(t,z)=M_{B}(z)\left\{1+\epsilon M_{1}(t,z)+O(\epsilon^{2})\right\}
\end{eqnarray}
with a small parameter $\epsilon$. The functions $S_{B}$,
$\eta_{B}$ and $M_{B}$ for the flat Friedmann solution are given by
\begin{equation}
 S_{B}(z) \propto (-z)^{-p}~, \qquad \eta_{B}(z) \propto z^{2}~,\qquad
  M_{B}(z) \propto (-z)^{2-3p}\label{eq:FF}
\end{equation}
with the constant $p$ defined as
\begin{equation}
 p \equiv \frac{2}{3(1+\alpha)}~.
\end{equation}
Note that the line element (\ref{eq:LE}) for the flat Friedmann solution
can be reduced to a form more familiar in cosmology through the
coordinate transformation $r\rightarrow\hat{r}\propto r^{1-p}$
 \cite{BJCarr:PRD62:2000}.

As seen from the behaviors such that $S_{B}\rightarrow 0$ and
$\eta_{B}\rightarrow\infty$ in the limit $z\rightarrow\infty$, 
a big crunch singularity appears at $t=0$ in the flat Friedmann
background. 
Therefore we hereafter consider time evolution of the perturbations during $t<0$
(i.e., $z<0$).
As was mentioned in Sec.~\ref{sec:intro}, in such a region there are two
 characteristic surfaces at which a constraint is imposed on the
 behavior of the perturbations.
One is the regular center located at $z=0^{-}$, and the other is the sonic
 point located at $z=z_{s}$ defined as a point at which the velocity of a
$z=\text{const}$ surface relative to the fluid is equal to the sound
speed, i.e., $V_{B}=-\sqrt{\alpha}$, where the function $V_{B}$ for the background flat
Friedmann solution is found to be 
\begin{equation}
 V_{B}(z) \propto (-z)^{1-p}~.
\end{equation}

Using the perturbation equations for $S_{1}$, $\eta_{1}$ and $M_{1}$, we can obtain the following single wave equation: 
\begin{equation}
 \Psi_{,uu}-\Psi_{,\zeta\zeta}+W(\zeta)\Psi_{,u}+F(\zeta)\Psi_{,\zeta}+U(\zeta)\Psi=0
\label{eq:wave}
\end{equation}
for the function $\Psi$ defined as
\begin{equation}
 \Psi(t,z) =S_{1}(t,z)-f(z)M_{1}(t,z)~, \label{eq:Psi}
\end{equation}
where the functions $W$, $F$, $U$ and $f$ are given in
Appendix~\ref{sec:functions}.
The new variables $u$ and $\zeta$ are related to the variables $t$ and
$z$ by 
\begin{eqnarray}
 &&u = \log(-t)+I(z)~, \qquad I(z) = \frac{1}{2(1-p)}\log{\left(1-x^{2}(z)\right)}~,\label{eq:u}\\
 && \zeta = \frac{1}{2(1-p)}\log{\frac{1-x(z)}{1+x(z)}}~\label{eq:zeta},
\end{eqnarray}
where the function $x$ is defined as
\begin{equation}
 x = -\frac{V_{B}}{\sqrt{\alpha}}~.\label{eq:x}
\end{equation}
The remaining perturbation equations yield the relations
\begin{eqnarray}
 M_{1}(t,z)&=&B_{1}(z)\dot{\Psi}(t,z)+B_{2}(z)\Psi'(t,z)+B_{3}(z)\Psi(t,z)~,\label{eq:M1}\\
\eta_{1}(t,z)&=&M_{1}(t,z)+\frac{1}{3(1-p)}M_{1}'(t,z)-3S_{1}(t,z)-\frac{1}{1-p}S_{1}'(t,z)~,\label{eq:eta1}
\end{eqnarray}
where the dot and the prime represent the partial derivative with
respect to $\log|t|$ and $\log|z|$, respectively, and 
the functions $B_{1}$, $B_{2}$ and $B_{3}$ are also given in
Appendix~\ref{sec:functions}.
The perturbations $S_{1}$, $\eta_{1}$ and $M_{1}$ are determined by the
solution $\Psi$ for the wave equation (\ref{eq:wave}) via
Eqs.~(\ref{eq:M1}), (\ref{eq:Psi}) and (\ref{eq:eta1}).

The wave equation (\ref{eq:wave}) allows us to consider 
the modes $\phi$ defined as 
\begin{equation}
 \Psi(t,z)=\phi(x,\omega)e^{i\omega (u+\zeta)}~\label{eq:Psib}
\end{equation}
with the spectral parameter $\omega$.
It is mathematically convenient to use the variable $x$,  
instead of $\zeta$, which 
can cover the whole region between the regular center and the sonic
point in the finite range $0 \leq x \leq 1$, irrespective of the
parameter value $\alpha$. Then the equation for $\phi$ is found to be 
\begin{equation}
 \phi_{,xx}+\frac{2i\omega-2(1-p)x-F}{(1-p)(1-x^{2})}\phi_{,x}-\frac{i\omega(F+W)+U}{(1-p)^{2}(1-x^{2})^{2}}\phi=0~.\label{eq:ODE}
\end{equation}

We will obtain the normal modes as the solutions for Eq.~(\ref{eq:ODE})
satisfying the boundary conditions such that $\phi$ is 
analytic both at the regular center $x=0$ and
at the sonic point $x=1$. (The same boundary conditions were required in
the numerical analysis done by \cite{HaradaT:PRD63:2001}.) 
The leading behavior of $\phi$ near $x=1$ is given by 
\begin{equation}
 \phi \simeq C_{1}(\omega)+C_{2}(\omega)(1-x)^{k}~,\label{eq:phi03}
\end{equation}
where $k=\{(1+i\omega)/(1-p)\}+1$.
The ratio $C_{2}/C_{1}$ will be uniquely determined by the requirement
of the analyticity of $\phi$ at $x=0$, and discrete eigenvalues
$\omega=\omega_{n}$ giving the normal modes
$\phi(x,\omega_{n})\equiv \phi_{n}(x)$ will be derived by the equation 
\begin{equation}
 C_{2}(\omega)=0~. \label{eq:BCS}
\end{equation}

Here we would like to note that there exists an
exact solution $\phi=\phi_{g}$ for Eq.~(\ref{eq:ODE}) written by
\begin{equation}
 \phi_{g} \propto \frac{x(1+x)^{3(1-\alpha)/(3\alpha+1)}}{2x+1+3\alpha}
\end{equation}
if the spectral parameter $\omega$ is equal to $\omega_{g}$ defined as
\begin{equation}
 \omega_{g}=\frac{1-\alpha}{1+\alpha}i~.
\end{equation}
This solution $\phi_{g}$ is one of the normal modes $\phi_{n}$ but corresponds to a
gauge mode due to an infinitesimal transformation of $t$.
In fact, all the perturbations $\eta_{1}$, $S_{1}$ and $M_{1}$ obtained
from $\phi_{g}$ are found to be independent of $z$.
Although the gauge mode is obviously unphysical, the presence of such an 
exact solution will be mathematically useful for checking the validity
of the analysis of Eq.~(\ref{eq:ODE}).

\section{Normal modes in the low pressure limit \label{sec:normal}}

Although the master equation (\ref{eq:ODE}) for the perturbations in the
flat Friedmann background is given by a simpler form (if compared with
the form for any other self-similar backgrounds), it is still a difficult task to
solve the eigenvalue problem, and in \cite{MitsudaE:PRD73:2006} the
absence of the unstable normal modes was clearly proven only in a
limited range of $\omega$.
Such a difficulty may be overcome if we consider the low pressure limit 
$\alpha\rightarrow0$, keeping the variable $x$ finite in the range
$0\leq x \leq 1$ and expanding the solution $\phi(x,\omega,\alpha)$ analytic at
$x=0$ as follows,
\begin{equation}
 \phi(x,\omega,\alpha)=\sum^{\infty}_{i=0}\alpha^{i}\phi^{(i)}(x,\omega)~.\label{eq:expansion}
\end{equation}

For the lowest-order solution $\phi^{(0)}$ we obtain the equation
 \begin{equation}
 \mathcal{L}\phi^{(0)}(x,\omega)=0~,\label{eq:ODE(0)}
\end{equation}
where the ordinary differential operator $\mathcal{L}$ is given by
\begin{eqnarray}
 \mathcal{L}&=&\frac{d^{2}}{dx^{2}}+\frac{3}{1-x^{2}}\left\{2i\omega+\frac{2x^{3}+3x^{2}+10x+3}{3x(1+2x)}\right\}\frac{d}{dx}\nonumber\\
&&-\frac{3}{x(1+2x)(1+x)(1-x^{2})}\left\{2i\omega(2x^{2}-2x-1)-\frac{2x^{4}+4x^{3}-x^{2}-2x-1}{x}\right\}~.\nonumber\\
\end{eqnarray}
We would like to emphasize that this limit does not mean to consider an exactly
pressureless fluid (i.e., a dust fluid) because the requirement of the
analyticity of $\phi$ at the sonic point $x=1$ is not missed.
The crucial point in this approach is that we can explicitly derive
general solutions for Eq.~(\ref{eq:ODE(0)}).
In particular, the solution $\phi^{(0)}$ 
satisfying the boundary condition at $x=0$ is written as
\begin{equation}
 \phi^{(0)}(x,\omega)=Z_{1}(x,\omega)-Z_{2}(x,\omega)~,\label{eq:Z0}
\end{equation}
where the functions $Z_{1}$ and $Z_{2}$ are the two independent solutions for
Eq.~(\ref{eq:ODE(0)}) and given by
\begin{eqnarray}
 Z_{1}(x,\omega)&=&\frac{(1+x)^{3}\left\{-6x^{2}\omega^{2}-4x(x^{2}-3x+1)i\omega+(1-x)^{4}\right\}}{x^{3}(1+2x)}~,\\
 Z_{2}(x,\omega)&=&\frac{(1-x)^{4+3i\omega}(1+x)^{1-3i\omega}\left\{x^{2}+2(1+i\omega)x+1\right\}}{x^{3}(1+2x)}~.
\end{eqnarray}

It is clear that this solution $\phi^{(0)}$ becomes analytic also at
the sonic point $x=1$ if the spectral parameter $\omega$ is equal to $\omega_{n}^{(0)}$ given by 
\begin{equation}
 \omega_{n}^{(0)}=\frac{4-n}{3}i~\label{eq:eigenvalue}
\end{equation}
with $n$ defined as non-negative integers. However, for $n=0$, $2$ and
$4$, the function $\phi^{(0)}$ turns out to vanish.
Hence, the values of $\omega_{n}^{(0)}$ are given only for
$n=1,3,5,6,\cdots$.
It can be easily found that the value of $\omega_{1}^{(0)}$ and the function 
$\phi^{(0)}(x, \omega_{1}^{(0)})$ are identical with the value of $\omega_{g}$ and
the gauge mode $\phi_{g}$ in the limit $\alpha\rightarrow 0$.
While this assures that $\omega_{n}^{(0)}$ represents the eigenvalues
approximately written in the limit $\alpha\rightarrow 0$, it may be
unclear how the condition (\ref{eq:BCS}) to obtain the eigenvalues
$\omega_{n}$ is used in this approach. 

To discuss this point, we consider the next order solution
$\phi^{(1)}(x,\omega)$ in Eq.~(\ref{eq:expansion}), using Eq.~(\ref{eq:Z0}).
The inhomogeneous ordinary differential equation
for the function $\phi^{(1)}$ is written as
\begin{equation}
 \mathcal{L}\phi^{(1)}(x,\omega)=J(x,\omega)~,\label{eq:ODE(1)}
\end{equation}
where the function $J$ is given in Appendix~\ref{sec:functions}.
Here we note that the imaginary part of $\omega_{n}^{(0)}$ is smaller than $4/3$.
Because we consider the solution $\phi^{(1)}$ analytic at
$x=0$ for $\omega$ nearly equal to $\omega_{n}^{(0)}$, from
Eq.~(\ref{eq:ODE(1)}) we obtain
\begin{equation}
 \phi^{(1)}(x,\omega)=Z_{2}(x,\omega)\int^{x}_{0}\frac{\phi^{(0)}(y,\omega)J(y,\omega)}{w(y,\omega)}dy+\phi^{(0)}(x,\omega)\int^{1}_{x}\frac{Z_{2}(y,\omega)J(y,\omega)}{w(y,\omega)}dy+a(\omega)\phi^{(0)}(x,\omega)~,\label{eq:phi(1)}
\end{equation}
where $a(\omega)$ is an arbitrary constant and the Wronskian $w$ of $\phi^{(0)}$
and $Z_{2}$ is given by
\begin{equation}
 w(x,\omega) = \phi^{(0)}Z_{2,x}-\phi_{,x}^{(0)}Z_{2}=-\frac{8(1-x)^{3(1+i\omega)}(1+x)^{3(1-i\omega)}\omega(\omega-2i)(3\omega-2i)(3\omega-4i)}{x^{3}(1+2x)^{2}}~.
\end{equation}

The eigenvalues $\omega_{n}(\alpha)$ giving the normal modes $\phi_{n}$
up to the first order of $\alpha$ will have the form
\begin{equation}
 \omega_{n}(\alpha) =
  \omega_{n}^{(0)}+\omega_{n}^{(1)}\alpha+O(\alpha^{2})~.
\label{eq:omegan}
\end{equation}
Because of this expansion of $\omega_{n}(\alpha)$, Eq.~(\ref{eq:expansion})
for the normal modes $\phi_{n}(\equiv \phi(x,\omega_{n}(\alpha),\alpha))$ can be rewritten to the form
\begin{equation}
\phi_{n}(x,\alpha)=\phi_{n}^{(0)}(x)+
\left(\frac{\partial \phi^{(0)}}{\partial \omega}(x,\omega^{(0)}_{n})\omega^{(1)}_{n}+\phi^{(1)}_{n}(x)\right)\alpha+O(\alpha^{2})~,\label{eq:phinex}
\end{equation}
where $\phi_{n}^{(i)}(x) \equiv \phi^{(i)}(x,\omega_{n}^{(0)})$.
In this expansion scheme, the function $\phi_{n}$ may be non-analytic at
$x=1$ owing to terms containing the logarithmic factor $\log(1-x)$
in $\partial \phi^{(0)}/\partial\omega (x, \omega_{n}^{(0)})$ and
$\phi_{n}^{(1)}$.
It is straightforward to derive such non-analytic terms, 
and we have
\begin{eqnarray}
 &&\frac{\partial \phi^{(0)}}{\partial\omega}(x,
  \omega_{n}^{(0)})=K_{n}(x)-3iZ_{2}(x,\omega_{n}^{(0)})\log(1-x)~,\label{eq:phin(0)}\\
 &&\phi_{n}^{(1)}(x) =L_{n}(x)-b_{n}Z_{2}(x,\omega_{n}^{(0)})\log(1-x)~,\label{eq:phin(1)}
\end{eqnarray}
where the functions $K_{n}$ and $L_{n}$ are analytic at $x=1$ and 
the coefficient $b_{n}$ is obtained from the term proportional to $(1-x)^{-1}$ in
the expansion of the function $Z_{1}J/w$ for $\omega=\omega_{n}^{(0)}$
around $x=1$. (Note that the function $Z_{1}J/w$ appears in the integrand
of the first integral in the right hand side of Eq.~(\ref{eq:phi(1)}) 
because of Eq.~(\ref{eq:Z0}).)
It is clear that the non-analyticity of $\phi_{n}$ at $x=1$ can be
removed if $\omega_{n}^{(1)}$ is chosen as follows, 
\begin{equation}
 \omega^{(1)}_{n} = \frac{b_{n}i}{3}~.\label{eq:omegan(1)}
\end{equation}

To estimate the coefficient $b_{n}$, we rewrite the function
$Z_{1}J/w$ for $\omega=\omega_{n}^{(0)}$ into the form  
\begin{equation}
 \frac{Z_{1}(x,\omega_{n}^{(0)})J(x,\omega_{n}^{(0)})}{w(x,\omega_{n}^{(0)})}=B(x,\omega_{n}^{(0)})\left\{(1-x)^{3-n}J_{1}(x,\omega_{n}^{(0)})+(1-x)^{-n}J_{2}(x,\omega_{n}^{(0)})+(1-x)^{-1}J_{3}(x,\omega_{n}^{(0)})\right\}~,\label{eq:integrand1}
\end{equation}
where the functions $J_{1}$, $J_{2}$ and $J_{3}$ are given in
Appendix~\ref{sec:functions} and the function $B$ is defined as
\begin{equation}
 B (x,\omega) \equiv \frac{Z_{1}(x,\omega)}{w(x,\omega)(1-x)^{-3(1+i\omega)}}~.
\end{equation}
Note that the functions $B$, $J_{1}$,
$J_{2}$ and $J_{3}$ for $\omega=\omega_{n}^{(0)}$ become finite at
$x=1$. From the form given by Eq.~(\ref{eq:integrand1}), we can easily find
\begin{equation}
 b_{n}=B(1,\omega_{n}^{(0)})J_{3}(1,\omega_{n}^{(0)})+\bar{b}_{n}~,\label{eq:bn}
\end{equation}
where $\bar{b}_{n}$ is given by
\begin{eqnarray}
 \bar{b}_{1}&=&
  B(1,\omega_{1}^{(0)})J_{2}(1,\omega_{1}^{(0)})~,\\
 \bar{b}_{3}&=& \left.\frac{1}{2}\left(B(x,\omega_{3}^{(0)})J_{2}(x,\omega_{3}^{(0)})\right)_{,xx}\right|_{x=1}~,
\end{eqnarray}
and for $n \geq 5$
\begin{equation}
 \bar{b}_{n}=\frac{(-1)^{n-4}}{(n-4)!}\left.\frac{d^{n-4}}{dx^{n-4}}\left(B(x,\omega_{n}^{(0)})J_{1}(x,\omega_{n}^{(0)})\right)\right|_{x=1}+\frac{(-1)^{n-1}}{(n-1)!}\left.\frac{d^{n-1}}{dx^{n-1}}\left(B(x,\omega_{n}^{(0)})J_{2}(x,\omega_{n}^{(0)})\right)\right|_{x=1}~.
\end{equation}
We can easily calculate the first term in the right hand side of
Eq.~(\ref{eq:bn}) to be
\begin{equation}
 B(1,\omega_{n}^{(0)})J_{3}(1,\omega_{n}^{(0)})=-2(n-1)~.\label{eq:BJ3}
\end{equation}
In addition we find that the value of $\bar{b}_{n}$ is given by the unified form
\begin{equation}
 \bar{b}_{n}=6(-1)^{n}\label{eq:barbn}
\end{equation}
for $n=1,3,5,6,\cdots$. 
Thus we arrive at the result
\begin{equation}
 \omega_{n}(\alpha)=\frac{4-n}{3}i+\left\{-\frac{2}{3}(n-1)+2(-1)^{n}\right\}\alpha
  i+O(\alpha^{2})\label{eq:omeganb}
\end{equation}
for the eigenvalue problem Eq.~(\ref{eq:BCS}) giving the normal modes $\phi_{n}$ written as
\begin{equation}
 \phi_{n}(x,\alpha) = \phi_{n}^{(0)}(x)+ \left[K_{n}(x)\left\{-\frac{2}{3}(n-1)+2(-1)^{n}\right\}i+L_{n}(x)\right]\alpha+O(\alpha^{2})~,\label{eq:phinb}
\end{equation}
where the functions $K_{n}$ and $L_{n}$ are explicitly given through
Eqs.~(\ref{eq:phin(0)}) and (\ref{eq:phin(1)}).
It should be noted that the value of $\omega_{1}$ given by the above
equation is identical with the value of $\omega_{g}$ up to the first
order of $\alpha$, as was expected.
Although the normal modes $\phi_{n}$ appear to have an ambiguity due to
the existence of the constant $a$ in Eq.~(\ref{eq:phi(1)}), it will
 be uniquely determined if we require the analyticity at $x=1$ up to the second order of $\alpha$. 
In fact, we can confirm that Eq.~(\ref{eq:phinb}) for $n=1$ is identical with the
gauge mode $\phi_{g}$ only for
$a=7-12\log2$.

The most important result is that the imaginary part of $\omega_{3}^{(0)}$
given by Eq.~(\ref{eq:eigenvalue}) is positive, namely, there exists one
unstable normal mode, at least for sufficiently small values of
$\alpha$. The proof concerning the absence of unstable normal modes
shown in \cite{MitsudaE:PRD73:2006} is not applicable to the normal mode $\phi_{3}$ with the small
growth rate $\text{Im}(\omega_{3}^{(0)}+\alpha\omega_{3}^{(1)})$ obtained
here. Hence the flat Friedmann solution does not act as an attractor in 
homogeneous collapse of a sufficiently low pressure perfect
fluid. However, the first order correction $\omega_{3}^{(1)}$ has a
negative imaginary part. This means that the growth rate 
of the unstable normal mode becomes smaller
as the value of $\alpha$ increases, and the self-similar behavior might be stable in
homogeneous collapse of a higher pressure perfect fluid such as the
radiation fluid ($\alpha=1/3$) and the stiff fluid ($\alpha=1$). 
This is an interesting problem to be
further studied in future works.  
Moreover, it is also interesting that there exists an infinite set of the stable
normal modes (i.e., $\phi_{n}$ for $n \geq 5$). 
In the next section, we will derive the density perturbation $\eta_{1}$
corresponding to the normal modes $\phi_{n}$ to see what configuration
of perturbed inhomogeneous fields can develop or decay with the lapse of time.

\section{Density perturbations\label{sec:interpretation}}

Let us denote the perturbed density $\eta_{1}$ corresponding to the
normal modes $\phi_{n}$ by $\eta_{1(n)}$, which is expanded with respect
to $\alpha$ as follows, 
\begin{equation}
 \eta_{1(n)}(t,x,\alpha) = \eta_{1(n)}^{(0)}(x)\exp{\{i\omega_{n}^{(0)}\log(-t)\}}+O(\alpha)~.
\end{equation}
Here we focus our attention on the leading term $\eta_{1(n)}^{(0)}$
depending on $x$.
Through Eqs.~(\ref{eq:Z0}), (\ref{eq:Psib}),
(\ref{eq:u}), (\ref{eq:zeta}), (\ref{eq:M1}), (\ref{eq:Psi}) and
(\ref{eq:eta1}) in the limit $\alpha \rightarrow 0$ with the eigenvalues $\omega=\omega_{n}^{(0)}$, we obtain 
\begin{eqnarray}
 \eta_{1(n)}^{(0)}(x)&=&\frac{2+n}{6x}\left[(1-x)^{n-2}\left\{3x^{2}+3(n-2)x+(n-1)(n-3)\right\}\right.\nonumber\\&&\left.\hspace{2cm}-(1+x)^{n-2}\left\{3x^{2}-3(n-2)x+(n-1)(n-3)\right\}\right]~.\label{eq:eta1n0}
\end{eqnarray}

\begin{figure}
\includegraphics[height=7cm]{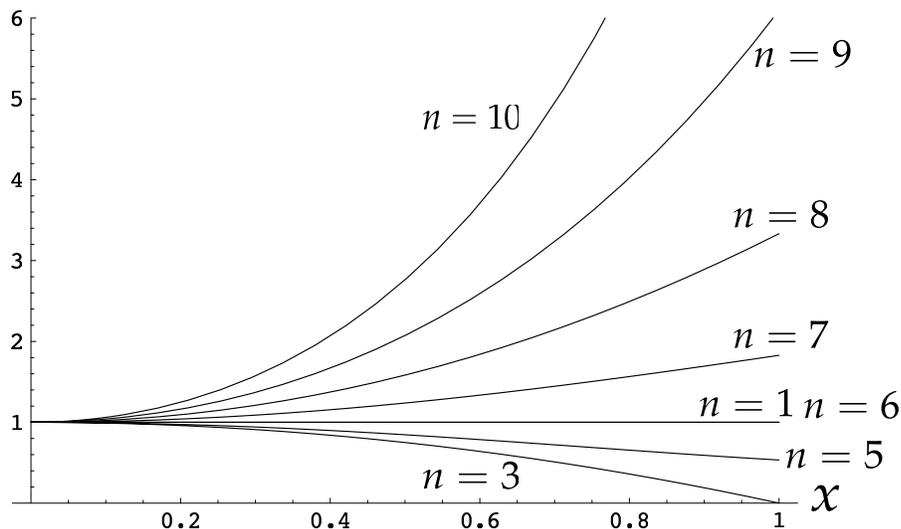}
\caption{Configuration of the density perturbation given by
 $\eta_{1(n)}^{(0)}(x)/\eta_{1(n)}^{(0)}(0)$.} 
\label{fg:eta1n0}
\end{figure}

In Fig.~\ref{fg:eta1n0}, we show the configuration of the density
perturbation 
$\eta_{1(n)}^{(0)}(x)$ normalized by its value at $x=0$, namely,  
\begin{equation}
 \eta_{1(n)}^{(0)}(0)=-\frac{(n-4)(n-2)n(n+2)}{3}~,\label{eq:eta1n00}
\end{equation}
for $n=1,3,5,6,7,8,9$ and $10$ in the range $0\leq x\leq 1$ .
It is shown in this figure that the normal mode
$\eta_{1(1)}^{(0)}$ corresponds to the gauge mode
$\eta_{g}=\text{const}$ at a given time as was mentioned in Sec.~\ref{sec:theory}.
In addition, it should be emphasized that the normal mode
$\eta_{1(6)}^{(0)}$, which is also constant at
any $x$, is a physical normal mode because the corresponding 
perturbation $M_{1}$ depends on $x$ in the range $0\leq x\leq 1$.

From Fig.~\ref{fg:eta1n0}, we note that the amplitude of all the stable
 normal modes (i.e., the normal modes for $n \geq 5$) at the sonic point
 $x=1$ remains non-zero and the amplitude of the normal modes for $n\geq 7$
rather increases towards the sonic point $x=1$ from the center $x=0$.
The ratio $\eta_{1(n)}^{(0)}(1)/\eta_{1(n)}^{(0)}(0)$ increases as $n$
 becomes larger and the decay rate $-\text{Im}(\omega_{n})$ given by
 Eq.~(\ref{eq:omeganb}) has the same tendency.
This implies that the density perturbation generated near the sonic
 point is rapidly carried away to the supersonic region $x>1$ by the
 background transonic flow and the growth of such a density perturbation in
 the subsonic region is prevented. 
It is remarkable that only for $n=3$, the value of $\eta_{1(n)}^{(0)}$
 vanishes at $x=1$. This seems to be the most favorable configuration of
 $\eta_{1}$ to allow the growth of the density perturbation due to the
 effect of its own self-gravitation.
Further, Eq.~(\ref{eq:omeganb}) clearly shows that the imaginary part of
 $\omega_{n}$ decreases as $\alpha$ increases. The dispersive effect due
 to the pressure against the self-gravitation can enhance the
 above-mentioned decay process of the perturbations.

In this section, we have focused our concern on the configuration of the
normal modes in the subsonic region $0\leq x \leq 1$. This is mainly
because the eigenvalue problem was set under the boundary conditions at
$x=0$ and $x=1$ and from the viewpoint of causality, any disturbances in
the supersonic region cannot affect the process in the subsonic region.
Further, as another reason, 
we would like to point out that the approximation $\alpha \simeq 0$ to
derive Eq.~(\ref{eq:omeganb}) becomes
mathematically unreliable in 
the region far away from the sonic point, i.e., $x \gg 1$.
If one tries to understand more global features of the normal modes by
analyzing Eq.~(\ref{eq:ODE}) in the limit $\alpha\rightarrow0$, 
the terms included in Eq.~(\ref{eq:ODE}) 
which can be negligible in the subsonic region but become important for $x \gg 1$ 
should be taken into consideration.

\section{Summary and Discussion \label{sec:summary}}

In this paper, we have studied the stability problem for the self-similar
behavior in homogeneous collapse of a perfect fluid with
the equation of state $P=\alpha\rho$, using the perturbation theory developed in
\cite{MitsudaE:PRD73:2006}. We have derived the single ordinary
differential equation (\ref{eq:ODE}) governing spherically symmetric
non-self-similar perturbations with the time dependence 
$\exp{\{i\omega\log{(-t)}\}}$ in the flat Friedmann background and set
up the eigenvalue problem to determine the value of the
spectral parameter $\omega$.
In the low pressure approximation $\alpha\rightarrow 0$, we have
succeeded in deriving explicitly an infinite set of the eigenvalues 
and the normal
modes given by Eqs.~(\ref{eq:omeganb})
and (\ref{eq:phinb}). Because one of such normal modes is an unstable
normal mode, we have 
concluded that non-self-similar inhomogeneous disturbances can develop 
in homogeneous collapse of a
sufficiently low pressure perfect fluid.

As was mentioned in Sec.~\ref{sec:intro}, the unstable normal mode
obtained in this paper was not found in the numerical analysis
\cite{HaradaT:PRD63:2001}, which has rather claimed that for $\alpha$ lying in the range $0<\alpha\lesssim 0.036$, the 
geometrical structure and the fluid motion at late stages in general
non-self-similar collapse of a perfect fluid (which is initially
homogeneous) can be well described by
the flat Friedmann solution. 
However, recalling that the unstable normal mode obtained in this paper
has the small growth rate (i.e., $\text{Im}(\omega_{3})$) less than $1/3$, we suggest that even if the 
geometrical structure and the fluid motion once become similar to those
of the flat Friedmann solution,  
non-self-similar disturbances become significant at much later stages which were missed in the
numerical simulation \cite{HaradaT:PRD63:2001}.
Although the general relativistic version of the Larson-Penston solution
was also suggested to act as an attractor in general inhomogeneous collapse 
by the results of the numerical simulation and the normal mode analysis
in \cite{HaradaT:PRD63:2001}, we claim that its stability should be also
confirmed in our analytical scheme. 

Because of the above-mentioned result of the numerical simulation
\cite{HaradaT:PRD63:2001}, the result obtained in this paper allows us
to interpret the flat Friedmann
solution as an intermediate attractor in
general non-self-similar perfect fluid collapse which starts from a
nearly homogeneous density profile, at
least for sufficiently small $\alpha$.
If the general relativistic Larson-Penston solution is confirmed to be
stable, a transition from the flat Friedmann stage to the general
relativistic Larson-Penston stage may occur in gravitational
collapse. (Such a transition was also mentioned in
\cite{HaradaT:PRD63:2001}.)
The so-called critical solution corresponding to
the threshold between the black hole formation and the complete 
dispersion of the fluid is the well-studied self-similar perfect fluid
solution (see e.g., \cite{NeilsenDW:CQG17:2000}) acting
as an intermediate attractor in general inhomogeneous collapse. 
It is interesting to note that there is a common feature between the flat Friedmann solution and
the critical solution, namely, the single unstable normal mode exists for these solutions.
It was found from the idea of the renormalization group in \cite{KoikeT:PRL74:1995, KoikeT:PRD59:1999}
that such a feature is essential to the scaling-law and the
universality observed in the critical phenomena.
Therefore what critical phenomena are relevant to the flat Friedmann
solution will be an interesting problem to be investigated in future works.

\appendix
\section{Functions in the perturbation equations\label{sec:functions}}

It is an easy task to calculate the functions involved in
Eqs.~(\ref{eq:wave}) and (\ref{eq:M1}) by using the formulae given in  
\cite{MitsudaE:PRD73:2006}.
The results are summarized as follows, 
\begin{eqnarray}
 W&=&
  \frac{(1+3\alpha)x^{3}+(7+9\alpha)x^{2}+(1+3\alpha)(7+9\alpha)x+(1+3\alpha)^{2}}{3(1+\alpha)x(1+3\alpha+x)}~,\\\label{eq:W}
 F &=&
  -\frac{6(1-\alpha)x^{4}-(1+3\alpha)(9\alpha-11)x^{3}+3(1+3\alpha)(5+7\alpha)x^{2}+(1+3\alpha)^{2}(13+3\alpha)x+3(1+3\alpha^{3})}{3(1+\alpha)x(1+3\alpha+x)(1+3\alpha+2x)}~,\label{eq:F}\\
 U &=& \frac{(1+3\alpha)(1-x^{2})\left\{2(\alpha-1)x^{4}+4(\alpha-1)(1+3\alpha)x^{3}-(1+3\alpha)(5\alpha-1)x^{2}+2(1+3\alpha)^{2}x+(1+3\alpha)^{3}\right\}}{3(1+\alpha)^{2}x^{2}(1+3\alpha+x)(1+3\alpha+2x)}~,\label{eq:U}\nonumber\\
f &=&\frac{2\alpha x-3\alpha-1}{3\alpha(1+3\alpha+2x)}~,\label{eq:f}\\
B_{1}&=&
  \frac{9\alpha(1+3\alpha+2x)^{2}}{8(1+3\alpha)(1+3\alpha+x)}~,\label{eq:T1}\\
  B_{2}&=&
  -\frac{9\alpha(1+x)(1+3\alpha+2x)^{2}}{8(1+3\alpha)x(1+3\alpha+x)}~,\label{eq:T2}\\
 B_{3}&=&
  \frac{9\alpha(1+3\alpha+2x)\left\{2(\alpha-1)x^{2}+(1+3\alpha)^{2}\right\}}{8(1+\alpha)(1+3\alpha)x(1+3\alpha+x)}~.\label{eq:T3}
\end{eqnarray}

If the functions involved in Eq.~(\ref{eq:ODE}) are expanded with
respect to $\alpha$ as follows, 
\begin{eqnarray}
 &&F+2(1-p)x =Q_{0}(x)+Q_{1}(x)\alpha+O(\alpha^{2})~,\\
 &&F+W=P_{0}(x)+P_{1}(x)\alpha+O(\alpha^{2})~,\\
 &&U=U_{0}(x)+U_{1}(x)\alpha+O(\alpha^{2})~,
\end{eqnarray}
 the inhomogeneous term $J$ in Eq.~(\ref{eq:ODE(1)}) is given by 
\begin{equation}
 J(x,\omega)=-\frac{1}{1-x^{2}}\left\{-6(2i\omega-Q_{0})-3Q_{1}\right\}\phi_{,x}^{(0)}+\frac{1}{(1-x^{2})^{2}}\left\{-36(i\omega
			 P_{0}+U_{0})+9(i\omega P_{1}+U_{1})\right\}\phi^{(0)}~,
\end{equation}
which leads us to the final result
\begin{eqnarray}
 J(x,\omega)&=&\frac{6}{(1-x)(1+x)^{2+3i\omega}x^{4}(1+2x)^{4}}\nonumber\\&&\times\left[(1+x)^{3+3i\omega}\left\{(x-1)^{3}(64x^{6}+122x^{5}+119x^{4}+74x^{3}+34x^{2}+16x+3)\right.\right.\nonumber\\&&-2(104x^{8}-102x^{7}-45x^{6}+4x^{5}-109x^{4}-43x^{3}+2x^{2}+7x+2)i\omega\nonumber\\&&\left.-2x(160x^{6}-122x^{5}-217x^{4}+67x^{3}+98x^{2}+28x+4)\omega^{2}+36x^{4}(1+2x)(3+4x)i\omega^{3}\right\}\nonumber\\&&+(1-x)^{3+3i\omega}\left\{(1+x)^{3}(64x^{6}+122x^{5}+119x^{4}+74x^{3}+34x^{2}+16x+3)\right.\nonumber\\&&+2(88x^{8}+420x^{7}+880x^{6}+1047x^{5}+821x^{4}+447x^{3}+155x^{2}+28x+2)i\omega\nonumber\\&&\left.\left.-2x(1+2x)(56x^{5}+238x^{4}+336x^{3}+225x^{2}+73x+8)\omega^{2}-24x^{2}(1+x)(1+2x)^{3}i\omega^{3}\right\}\right]~.
\end{eqnarray}
In addition, the functions $J_{1}$, $J_{2}$ and $J_{3}$ in
Eq.~(\ref{eq:integrand1}) are given by
\begin{eqnarray}
 J_{1}(x,\omega)&=&-\frac{6(1+x)}{x^{4}(1+2x)^{4}}(64x^{6}+122x^{5}+119x^{4}+74x^{3}+34x^{2}+16x+3)~,\\
 J_{2}(x,\omega)&=&\frac{6i\omega(1+x)}{x^{4}(1+2x)^{4}}\left\{-2(104x^{8}-102x^{7}-45x^{6}+4x^{5}-109x^{4}-43x^{3}+2x^{2}+7x+2)\right.\nonumber\\&&\left.+2x(160x^{6}-122x^{5}-217x^{4}+67x^{3}+98x^{2}+28x+4)i\omega+36x^{4}(1+2x)(3+4x)\omega^{2}\right\}~,\\
 J_{3}(x,\omega)&=&\frac{6}{(1+x)^{2+3i\omega}x^{4}(1+2x)^{4}}\nonumber\\&&\times\left\{(1+x)^{3}(64x^{6}+122x^{5}+119x^{4}+74x^{3}+34x^{2}+16x+3)\right.\nonumber\\&&+2(88x^{8}+420x^{7}+880x^{6}+1047x^{5}+821x^{4}+447x^{3}+155x^{2}+28x+2)i\omega\nonumber\\&&\left.-2x(1+2x)(56x^{5}+238x^{4}+336x^{3}+225x^{2}+73x+8)\omega^{2}-24x^{2}(1+x)(1+2x)^{3}i\omega^{3}\right\}~.
\end{eqnarray}

\bibliography{ref}

\end{document}